\documentclass[aps,prc,prcsuperscriptaddress,nofootinbib,11pt]{revtex4}

\usepackage{graphicx}  
\usepackage{bm}        
\usepackage{amssymb}   
\usepackage{lineno}

\usepackage[bookmarks]{hyperref}

\usepackage{color}
\definecolor{dgreen}{cmyk}{1.,0.,1.,0.2}        
\definecolor{orange}{cmyk}{0.,0.353,1.,0.}    

\def\snn{\mbox{$\sqrt{s_{_{\rm NN}}}$}}
\newcommand{ \be }{\begin{eqnarray}}
\newcommand{ \ee }{\end{eqnarray}}

\newcommand{ \sss }{\scriptscriptstyle}

\newcommand{ \mean }[1]{\left\langle #1 \right\rangle}

\newcommand{ \psirp }{\Psi_{\rm \sss RP}}
\newcommand{ \psiEP }{\psi_{2, EP}}
\newcommand{ \phia }{\phi_{\alpha}}
\newcommand{ \phib }{\phi_{\beta}}
\newcommand{ \phicl }{\phi_{\rm clust}}

\newcommand{ \Psisp }{\Psi_{1,\rm {\sss SP}}}
\newcommand{ \psisp }{\psi_{1,\rm {\sss SP}}}
\newcommand{ \psipp }{\Psi_{2,\rm {\sss PP}}}
\newcommand{ \psib }{\Psi_{2,{\rm \sss B}}}

\newcommand{\pt}{p_T}

\newcommand{ \CME}{{\rm \sss CME}}
\newcommand{ \PP}{{\rm \sss PP}}
\newcommand{ \BG}{{\rm \sss BG}}
\newcommand{ \dgamma }{\Delta\gamma}
\newcommand{ \dgBG }{\Delta\gamma^\BG}
\newcommand{ \dgCME }{\Delta\gamma^\CME}

\newcommand{ \dgcme }{\Delta\gamma^{\rm {\sss CME}}}

\newcommand{ \mvppsq }{\mean{v^2_{\rm 2, \PP}}}

\newcommand{ \vtwo }{v_2\{2\}}
\newcommand{ \vtwob }{v_2\{\psib \}}
\newcommand{ \vtwosp }{v_2\{\psisp\}}
\newcommand{ \vtwoSP }{v_2\{\Psisp\}}
\newcommand{ \vfour }{v_2\{4\}}

\newcommand{ \dgrsp}{\left( \dgamma/v_2 \right)_{\rm {\sss SP}}}

\newcommand{ \dgrc}{\left( \dgamma/v_2 \right)_{\rm c}}

\begin{document}




\title{Estimate of the CME signal in heavy-ion collisions
from measurements relative to
the  participant and spectator flow planes}

\author{Sergei A. Voloshin}
\affiliation{Wayne State University, 666 W. Hancock, Detroit, MI 48201}

\begin{abstract}
An interpretation of the charge dependent correlations sensitive to
the Chiral Magnetic Effect (CME) -- the separation of the electric
charges along the system magnetic field (across the reaction plane) --
is ambiguous due to a possible large background (non-CME) effects. The
background contribution is proportional to the elliptic flow $v_2$; it
is the largest in measurements relative to the participant plane, and
is smaller in measurements relative to the flow plane determined by
spectators, where the CME signal, on opposite, is likely larger. In
this note I discuss a possible strategy for corresponding experimental
measurements, and list and evaluate different assumptions related to
this approach.
\end{abstract}


\maketitle



The search for the chiral magnetic effect
(CME)~\cite{Kharzeev:2004ey,Fukushima:2008xe} -- the separation of the
electrical charges along the magnetic field in a chirally asymmetric
medium -- is a very active topic in the field of heavy ion collisions
for more than 10 years (for recent reviews,
see~\cite{Kharzeev:2015znc,Skokov:2016yrj}). The CME states that
particles originating from the same ``P-odd domain'' are
preferentially emitted either along or opposite to the magnetic field
direction depending on the particle charge.  As only a few particles
(originating from the same domain) are correlated, the signal is
expected to be small and one has to suppress other charge-dependent
correlations, such as due to the resonance decays, charge ordering in
jets, etc.. The so-called ``gamma'' correlator suggested in
Ref.~\cite{Voloshin:2004vk} was designed to do just that -- to
suppress non-CME correlation at least by a factor $\sim v_2$ --
the typical value of elliptic flow.
\be 
\gamma_{\alpha \beta} = \mean{\cos(\phia + \phib-2\Psi)}=
\mean{\sin(\phia-\Psi)\sin(\phib-\Psi)} -
\mean{\cos(\phia-\Psi)\cos(\phib-\Psi)} ,
\label{egamma}
\ee 
where $\phia$ and $\phib$ are the azimuthal angles of two charged
particles, $\alpha$ and $\beta$ taking values ``$+$'' or ``$-$''
denote the charge. $\Psi$ denotes the azimuth of the plane across
which the charge separation is measured. For measurements relative to
the reaction plane (perpendicular to the direction of the magnetic
field) only ``sin-sin'' term has contribution from the CME, while all
other non-CME sources contribute to both, ``sin-sin'' and ``cos-cos''
terms and thus largely cancel. The remaining difference between
``in-plane'' (``cos-cos'') and ``out-of-plane'' (sin-sin'')
correlations constitutes the background to the CME measurements via
gamma correlator. The background is zero in case of no elliptic flow
present in the system.

The experimental
measurements~\cite{Abelev:2009ac,Abelev:2009ad,Abelev:2012pa} are in
qualitative agreement with the theoretical expectations, but a
reliable separation of the CME signal from background effects is still
missing. As already mentioned, the background correlations depend on
the magnitude of elliptic flow and as such are largest in the
measurements performed relative to the so-called {\em participant
  plane}, and should be smaller in measurements relative to the
spectator flow plane. On opposite, the CME signal, driven by the
magnetic field, is likely larger in measurements relative to the
spectator plane, as the magnetic field is mostly determined by
spectator protons.  This idea was recently and independently used in
Ref.~\cite{Xu:2017qfs}, where the authors attempted to estimate the
CME signal from the existing measurements as well as make prediction
for the future isobar collision measurements at RHIC.  In this short
note I discuss an evaluation of the CME signal based on the same
general idea from a different perspective. In particular, I discuss in
detail the role of flow fluctuations in measurements relative to
different flow planes and by different methods, as well as explicitly
list different assumptions required in this approach, some of which
are more important than others.

I start with more definitions and recalling the derivation of the
background contribution to the gamma correlator. The correlator defined in
Eq.\ref{egamma} includes contributions from charge independent effect
(e.g. dipole flow). These are poorly known and not very important
for the CME search. Due to this we will discuss later only the charge
dependent part
\be \dgamma=\gamma_{\rm opposite} -\gamma_{same}. 
\ee
As both, the CME signal and the background correlations are small, one can
safely {\em assume} that
\be
\dgamma=\dgBG + \dgCME, 
\label{edec}
\ee 
neglecting, in principle possible, interplay between the two effects.
The background contribution to $\dgamma$ very generally can be
described as that due to ``flowing clusters''~\cite{Voloshin:2004vk}, when
both particles, $\alpha$ and $\beta$ belong to the same ``cluster'':  
\be
\dgBG=\Delta
\mean{\cos(\alpha+\beta-2\Psi)}=
\Delta \mean{\cos(\alpha+\beta-2\phicl)\,
  \cos(2\phicl-2\Psi)}_{\alpha,\beta\,\in\,{\rm clust}},
\label{edgBG}
\ee 
where to simplify notations we started to use symbols $\alpha$ and $\beta$
instead of $\phia$ and $\phib$.  Note that the mean of the product of
two cosines in general does not factorize. The mean can be non-zero
either in the case of non-zero elliptic flow of clusters, $
\mean{\cos(2\phicl-2\Psi)}$, or due to the fact that the ``kinematic''
factor $\mean{\cos(\alpha+\beta-2\phicl)}$ varies with the cluster
emission angle, or both (as in the case of the so-called ``local
charge conservation'' background~\cite{Schlichting:2010qia}).  The
first assumption about background is
\newline
(A1a) $\dgBG\propto \tilde{v}_{\rm 2;clust}$, where we used ``tilde''
to denote the fact that there might be no factorization in
Eq.~\ref{edgBG} in which case this flow coefficient also accounts for
the emission angle dependence of the ``kinematic factor''.
\newline
The assumption (A1a) by itself is not very useful without
further assumption on $\tilde{v}_{\rm 2;clust}$: 
\newline
(A1b) $\tilde{v}_{\rm 2;clust}\propto v_2$, where $v_2$ is the average
(over some rapidity and $\pt$ ranges) elliptic flow of charged
particles.  One can combine (A1a) and (A1b) into one assumption
\newline
(A1) $\dgBG= b\, v_2$, where by $b$ we denote the proportionality
constant. 
\newline
This is the assumption employed almost in any attempt to disentangle
background effects from the CME signal, e.g. used by ALICE and CMS
Collaborations~\cite{Acharya:2017fau,Sirunyan:2017quh} in the
estimates of the CME signal with the Event Shape Engineering
technique~\cite{Schukraft:2012ah}.  Reiterate, that (A1a) assumes {\em
  linear} dependence of the background contribution to $\dgamma$ on
$\tilde{v}_{\rm 2;clust}$, and (A1b) assumes the proportionality of
the latter to the elliptic flow of charged particles.

Due to the initial state fluctuations, the elliptic flow, as well as
the elliptic flow fluctuations, measured relative to different flow
symmetry planes, are different.  Then it becomes convenient to modify
the correlator, namely consider, $(\dgamma/v_2)$ with $v_2$ calculated
in the same way as the $\gamma$ itself:
\be
(\dgamma/v_2)
=\frac{\mean{\cos(\alpha+\beta-2\psi)}}{\mean{\cos(2a-2\psi)}},
\label{eratio}
\ee
where, for simplicity, we omit the sign $\Delta$ in the numerator
(here and everywhere below in the expressions involving particles
$\alpha$ and $\beta$ we assume taking the difference between opposite
and same charge combinations); $a$ stands for the same set of
particles as $\alpha$ and $\beta$, and the average is performed
inclusively of all charges.  In the denominator mesasuremnt it is
assumed that the non-flow contribution is eliminated/suppressed.  Note
that the calculations of this ratio does not involve any explicit
correction for the so-called reaction plane resolution. To emphasize
this, here and below we denote all the flow planes that include
statistical fluctuations (due to finite number of particles used for
their determination~\cite{Voloshin:2007pc}) with lower case $\psi$,
and the angles that do not include statistical fluctuations (depend
only on specific initial configuration) with upper case $\Psi$.

An important feature of the ratio Eq.~\ref{eratio} is that in the case
of zero CME-signal (pure background) this ratio is the same
irrespectively of what is used for the $\psi$ and how strongly (or
weakly) elliptic flow fluctuates relative to this plane.  Namely, in
the no-CME case this ratio equals $b$ -- the proportionality
coefficient in the assumption A1.  For example, if instead of $\psi$
the azimuthal angle of a particle $c$ is used, this ratio equals
\be 
\dgrc=\frac{\mean{\cos(\alpha+\beta-2c)}}{\mean{\cos(2a-2c)}}
=\frac{b\mvppsq}{\mvppsq}=b,
\ee 
where again for shorter notations we use the particle symbol to denote
the particle azimuthal angle.  
Note, that this
case corresponds to elliptic flow measured with respect to the
participant plane, and $\mvppsq = v_2^2\{2\}$.  
For simplicity we also assume that
the flow of both particles, $a$ and $c$ are the same.

Instead of $\psi$ in Eq.~\ref{eratio} one
can use the ``event plane'' angle $\psiEP$ (the azimuth of the flow
vector in another subevent), or, what is more relevant for this
discussion, the spectator flow angle $\psisp$
\be
\dgrsp = \frac{\mean{\cos(\alpha+\beta-2\psisp)}}
             {\mean{\cos(2a-2\psisp)}} .
\ee
Under the ``background scenario'' all these ratios equal one to
another. If two different measurements yield different ratios this
would immediately indicate a contribution different from that of
``background'', namely, the CME. 
Note that in calculations of the
denominators (flow with respect to different angles) ``non-flow''
contribution should be eliminated/suppressed (e.g. by imposing a
rapidity gap in measurements or by any other technique).  
If two ratios differ one can try to estimate the CME
signal. This will rely on further assumptions, but as we discuss
below, the requirement to ``accuracy'' of those is lower.

In the case of a non-zero CME signal the ratios  Eq.~\ref{eratio}
calculated relative to different angles can be different. For
concreteness let us consider the double ratio 
\be
\frac{\dgrsp}{\dgrc} = \frac{\mean{\cos(\alpha+\beta-2\psisp)}
             /\mean{\cos(2a-2\psisp)}}
     {\mean{\cos(\alpha+\beta-2c)}/\mvppsq}, 
\label{edouble}
\ee
where as above we assume the same elliptic flow of particles $a$ and
$c$. Recall also, that the particles $a$ and $c$ flowing in the
participant plane. For the discussion of the CME contribution, we
introduce the angle $\psib$ for the orientation of the plane
perpendicular to the magnetic field (across which the maximum charge
separation occurs).  This angle is not measurable, and we will need to
make further assumptions below to relate the obtained expressions to the
experimental measurements. Then, decomposing the correlators in
background and the signal parts similarly to Eq.~\ref{edec}
\be
\mean{\cos(\alpha+\beta-2c)} = 
\mean{\cos(\alpha+\beta-2c)}^\BG
+\mean{\cos(\alpha+\beta-2c)}^\CME=b \mvppsq + \dgcme \vtwob,
\ee 
where $\dgcme = \mean{\cos(\alpha+\beta-2\psib)}^\CME$ and
$\vtwob=\mean{\cos(2c-2\psib)}$.  In a similar way 
\be
\mean{\cos(\alpha+\beta-2\psisp)} = b\, \mean{\cos(2a-2\psisp)}
  + \dgcme
  \mean{\cos(2\psib-2\psisp)}
\ee
Combining everything together 
\be
\frac{\dgrsp}{\dgrc}=1+f_\PP^\CME \left(
\frac{ \mean{\cos(2\psib-2\psisp)} \mvppsq   }{
  \mean{\cos(2a-2\psisp)} \vtwob }-1
\right)
\label{edr}
\ee
where 
\be
f_\PP^\CME =\frac{\mean{\cos(\alpha+\beta-2c)}^\CME}
{\mean{\cos(\alpha+\beta-2c)}}
\ee
is the fraction of the CME signal in 3-particle correlator measured
relative to the second harmonic participant plane. The angle $\psisp$
fluctuates around the spectator plane $\Psisp$, but one can see that
in the expression Eq,~\ref{edr} the corresponding event plane
resolution factors cancel out and
\be
\frac{\dgrsp}{\dgrc}=1+f_\PP^\CME \left(
\frac{ \mean{\cos(2\psib-2\Psisp)} \mvppsq   }{
  \vtwoSP \vtwob }-1
\right),
\label{edr2}
\ee
where $\vtwoSP= \mean{\cos(2a-2\Psisp)}$.

To proceed further one has to make assumptions about the relative
orientations of three angles, $\psipp$, $\Psisp$ and $\psib$. We
discuss a few ``reasonable'' scenarios below.  First, it is
instructive to compare the centrality dependence of $\vtwo$, $\vfour$,
and $\vtwoSP$~\cite{Wang:2005ab}. Recall also that to a good
approximation (exact in the so-called Gaussian model of eccentricity
fluctuations~\cite{Voloshin:2007pc}), $\vfour$ measures the flow
relative to the true reaction plane. Experimentally~\cite{Wang:2005ab}
in midcentral collisions, centrality $\sim40-50\%$, $\vtwoSP$ is very
close to $\vfour$; it is much closer to $\vtwo$ in central, $<10\%$,
collisions. A possible interpretation of that would be that the
spectator plane is close to the reaction plane in midcentral
collisions and close to the participant plane in central collisions.

Having this in mind, one of the assumption would be  
\newline
(A2) in midcentral collisions, both, the spectator plane and the magnetic
field plane, coincide with the reaction plane.    
In this case
\be
\frac{\dgrsp}{\dgrc} = 1+f_\PP^\CME \left(
\frac{\mvppsq }{(\vtwoSP)^2}-1
\right)
\label{erf}
\ee
Note that  this relation really requires only coincidence of $\Psisp$
and $\psib$, not necessarily coincidence with $\psirp$. Then 
Eq.~\ref{erf} is also
true even if
\newline
(A3) in central collision $\psib$ deviates from $\psirp$ but 
coincides with $\Psisp$.

It is interesting that one has the same relation event under
quite differrent assumption that
\newline
(A4) in central collision the spectator plane coincides with
participant plane but, $\psib$ coincides with $\psirp$. In this case
\be
\frac{\vtwob}{ \mean{\cos(2\psib-2\Psisp)}}=
  \vtwoSP 
\ee
and one again arrives to Eq.~\ref{erf}.

Although in general it is difficult to get the exact value of the
expression in parenthesis in Eq.~\ref{edr}, based on the above
assumptions (A2)-(A4), and having in mind that experimentally
$\vtwo$ is larger than $\vtwosp$ by about $\sim 15\%$, one can conclude
that for an estimate of the CME fractional contribution to the
gamma correlator $ f_\PP^\CME $ at the level $\sim5\%$, the
ratio Eq.~\ref{eratio} should be measured with an accuracy better than
1\%. 

Finally we make two short remarks on the experimental selection of the
angles $\psisp$ and its relation to $\psib$.  Experimentally $\psisp$
is usually measured with zero degree calorimeters (ZDC), most often
capturing only neutrons. Then (a) an additional decorrelations between
$\psisp$ and $\psib$ can arise due to difference in plane determined
by spectator neutrons and spectator protons.  If two ZDC are used,
then (b) the result might depend on how the angles from two detectors are
used in the analysis. For example using only one of ZDCs might yield
$\psisp$ which is stronger correlated with the participant plane,
while combining two angle might eliminate this bias.

In conclusion, we show that measuring the ratios Eq.\ref{eratio}
relative to the participant and spectator planes can be used to
determine the fraction of the CME signal in the gamma correlator
measurements. If the double ratio, Eq.~\ref{edouble},
 deviated from unity it will indicate a
non-zero CME contribution that can be further quantified under
reasonable assumptions. On order to measure the fractional CME signal
at the level of $\sim 5\%$ one would need to measure the ratio
Eq.\ref{edouble} free from non-flow effect at the level of about 1\%.

\vspace{1mm}
{\bf{Acknowledgements.}}
The authors appreciate early discussions of this question with
Dr. I.~Selyuzhenkov.
This material is based upon work supported by the U.S. Department of 
Energy Office of Science, Office of Nuclear Physics under Award 
Number DE-FG02-92ER-40713.



\begin{thebibliography}{99}

\itemsep=0cm

\bibitem{Kharzeev:2004ey} 
  D.~Kharzeev,
  ``Parity violation in hot QCD: Why it can happen, and how to look for it,''
  Phys.\ Lett.\ B {\bf 633}, 260 (2006)
  doi:10.1016/j.physletb.2005.11.075
  [hep-ph/0406125].

\bibitem{Fukushima:2008xe} 
  K.~Fukushima, D.~E.~Kharzeev and H.~J.~Warringa,
  ``The Chiral Magnetic Effect,''
  Phys.\ Rev.\ D {\bf 78}, 074033 (2008)
  doi:10.1103/PhysRevD.78.074033
  [arXiv:0808.3382 [hep-ph]].

\bibitem{Kharzeev:2015znc} D.~E.~Kharzeev, J.~Liao, S.~A.~Voloshin and
  G.~Wang, 
 ``Chiral magnetic and vortical effects in high-energy
  nuclear collisions<E2><80><94>A status report,''
  Prog.\ Part.\ Nucl.\ Phys.\ {\bf 88}, 1 (2016)
  doi:10.1016/j.ppnp.2016.01.001 

\bibitem{Skokov:2016yrj} 
  V.~Koch, S.~Schlichting, V.~Skokov, P.~Sorensen, J.~Thomas, S.~Voloshin, G.~Wang and H.~U.~Yee,
  ``Status of the chiral magnetic effect and collisions of isobars,''
  Chin.\ Phys.\ C {\bf 41}, no. 7, 072001 (2017)
  doi:10.1088/1674-1137/41/7/072001
  [arXiv:1608.00982 [nucl-th]].

\bibitem{Voloshin:2004vk} 
 S.~A.~Voloshin,
  ``Parity violation in hot QCD: How to detect it,''
  Phys.\ Rev.\ C {\bf 70}, 057901 (2004)
  doi:10.1103/PhysRevC.70.057901
  [hep-ph/0406311].
  
\bibitem{Abelev:2009ac} 
  B.~I.~Abelev {\it et al.} [STAR Collaboration],
 ``Azimuthal Charged-Particle Correlations and Possible Local Strong Parity Violation,''
  Phys.\ Rev.\ Lett.\  {\bf 103}, 251601 (2009)
  doi:10.1103/PhysRevLett.103.251601
  [arXiv:0909.1739 [nucl-ex]].
  

\bibitem{Abelev:2009ad} 
  B.~I.~Abelev {\it et al.} [STAR Collaboration],
  ``Observation of charge-dependent azimuthal correlations and possible local strong parity violation in heavy ion collisions,''
  Phys.\ Rev.\ C {\bf 81}, 054908 (2010)
  doi:10.1103/PhysRevC.81.054908
  [arXiv:0909.1717 [nucl-ex]].


\bibitem{Abelev:2012pa} 
  B.~Abelev {\it et al.} [ALICE Collaboration],
  ``Charge separation relative to the reaction plane in Pb-Pb collisions at $\sqrt{s_{NN}}= 2.76$ TeV,''
  Phys.\ Rev.\ Lett.\  {\bf 110}, no. 1, 012301 (2013)
  doi:10.1103/PhysRevLett.110.012301
  [arXiv:1207.0900 [nucl-ex]].
  
\bibitem{Xu:2017qfs} 
  H.~j.~Xu, J.~Zhao, X.~Wang, H.~Li, Z.~W.~Lin, C.~Shen and F.~Wang,
  ``Varying the chiral magnetic effect relative to flow in a single nucleus-nucleus collision,''
  arXiv:1710.07265 [nucl-th].


\bibitem{Schlichting:2010qia} 
  S.~Schlichting and S.~Pratt,
  ``Charge conservation at energies available at the BNL Relativistic Heavy Ion Collider and contributions to local parity violation observables,''
  Phys.\ Rev.\ C {\bf 83}, 014913 (2011)
  doi:10.1103/PhysRevC.83.014913
  [arXiv:1009.4283 [nucl-th]].


\bibitem{Acharya:2017fau} 
  S.~Acharya {\it et al.} [ALICE Collaboration],
  ``Constraining the magnitude of the Chiral Magnetic Effect with Event Shape Engineering in Pb-Pb collisions at $\sqrt{s_\mathrm{NN}}$ = 2.76 TeV,''
  Phys.\ Lett.\ B {\bf 777}, 151 (2018)
  doi:10.1016/j.physletb.2017.12.021
  [arXiv:1709.04723 [nucl-ex]].

\bibitem{Sirunyan:2017quh} 
  A.~M.~Sirunyan {\it et al.} [CMS Collaboration],
  ``Constraints on the chiral magnetic effect using charge-dependent azimuthal correlations in $p\mathrm{Pb}$ and PbPb collisions at the CERN Large Hadron Collider,''
  Phys.\ Rev.\ C {\bf 97}, no. 4, 044912 (2018)
  doi:10.1103/PhysRevC.97.044912
  [arXiv:1708.01602 [nucl-ex]].

\bibitem{Schukraft:2012ah} 
  J.~Schukraft, A.~Timmins and S.~A.~Voloshin,
  ``Ultra-relativistic nuclear collisions: event shape engineering,''
  Phys.\ Lett.\ B {\bf 719}, 394 (2013)
  doi:10.1016/j.physletb.2013.01.045
  [arXiv:1208.4563 [nucl-ex]].


\bibitem{Voloshin:2007pc} 
  S.~A.~Voloshin, A.~M.~Poskanzer, A.~Tang and G.~Wang,
  ``Elliptic flow in the Gaussian model of eccentricity fluctuations,''
  Phys.\ Lett.\ B {\bf 659}, 537 (2008)
  doi:10.1016/j.physletb.2007.11.043
  [arXiv:0708.0800 [nucl-th]].

\bibitem{Wang:2005ab} 
  G.~Wang [STAR Collaboration],
  ``Anisotropic flow in Au Au and Cu Cu at 62-GeV and 200-GeV,''
  Nucl.\ Phys.\ A {\bf 774}, 515 (2006)
  doi:10.1016/j.nuclphysa.2006.06.017
  [nucl-ex/0510034].
\end{thebibliography}
\end{document}